\documentclass[12pt]{iopart}

\usepackage[FIGTOPCAP,small]{subfigure}
\usepackage{graphicx}
\usepackage{dcolumn}
\usepackage{bm}

\usepackage{xcolor}

\newcommand{\bra}[1]{\left< #1 \right\vert}
\newcommand{\ket}[1]{\left\vert #1 \right>}
\newcommand{\pare}[1]{\left( #1 \right)}
\newcommand{\abs}[1]{\left\vert #1 \right\vert}
\newcommand{\cor}[1]{\left[ #1 \right]}

\newcommand{\llav}[1]{\left\lbrace #1 \right\rbrace}

\begin{document}

\title[]{Role of the spectral shape of quantum correlations in two-photon virtual-state spectroscopy}
\author{R de J Le\'on-Montiel,$^1$ J Svozil\'{i}k,$^{1,2}$ L J Salazar-Serrano,$^{1,3}$ and Juan P Torres$^{1,4}$}
\address{$^1$ICFO-Institut de Ciencies Fotoniques, Mediterranean
Technology Park, 08860 Castelldefels (Barcelona), Spain}
\address{$^2$RCPTM, Joint Laboratory of Optics PU and IP AS CR, 17. listopadu 12, 77146 Olomouc, Czech Republic}
\address{$^3$Physics Department, Universidad de los Andes, A.A. 4976, Bogot\'{a} D.C., Colombia}
\address{$^4$Department of Signal Theory and Communications,
Campus Nord D3, Universitat Politecnica de Catalunya, 08034
Barcelona, Spain}
\ead{roberto.leon@icfo.es}

\begin{abstract}
It is controversial what is the true role of entanglement in
two-photon virtual-state spectroscopy [Saleh  \emph{et al},
\emph{Phys. Rev. Lett.} \textbf{80}, 3483, 1998], a two-photon
absorption spectroscopic technique that can retrieve information
about the energy level structure of an atom or a molecule. The
consideration of closely related techniques, such as
multidimensional pump-probe spectroscopy [Roslyak \emph{et al},
\emph{Phys. Rev. A} \textbf{79}, 063409, 2009] suggests that
spectroscopic information might also be retrieved by using
uncorrelated pairs of photons. Here we show that this is not the
case. In the two-photon absorption process, the ability to obtain
information about the energy level structure of a medium depends
on the spectral shape of existing temporal (frequency)
correlations between the absorbed photons. In fact, it is a
combination of both, the presence of frequency correlations
(entanglement) and its specific spectral shape, which makes the
realization of two-photon virtual-state spectroscopy possible.
This result helps for selecting the type of two-photon source that
needs to be used in order to experimentally perform the two-photon
virtual-state spectroscopy technique.
\end{abstract}
\submitto{\NJP}
\maketitle
\section{Introduction}
The process of two-photon absorption (TPA), the
light-induced transition between two energy levels of a medium
mediated by the absorption of two photons, is a building block of
some technologies aimed at probing the structure of atoms and
molecules, such as two-photon microscopy \cite{denk1990} and
two-photon spectroscopy \cite{hopfield1965}. In particular,
nonlinear two-photon spectroscopy has become an invaluable tool
\cite{mukamel_book}, where the capability of TPA is exploited to
obtain information about a sample that would not be accessible
otherwise.

With the advent of light sources that generate entangled photon
pairs \cite{mandel}, new phenomena in TPA processes have been
unveiled. The linear dependence of the two-photon absorption rate
on the photon flux \cite{juha1990}, two-photon induced
transparency \cite{fei1997}, virtual-state spectroscopy
\cite{teich1998,kojima2004} and the selectivity of double-exciton
states of chromophore aggregates \cite{mukamel2012} are effects
that have been attributed to the presence of entanglement. However,
the link between entanglement and the new effect observed
is sometimes blur. So, might not be the ultimate cause of some of these
effects an accompanying characteristic unrelated to its entangled
nature? This is the case of certain effects that,
when first described, were attributed to the existence of
frequency entanglement between pairs of photons. For instance,
Nasr \emph{et al.} \cite{nasr2003} demonstrated a new scheme, based on
entanglement, to erase effects due to second-order chromatic
dispersion in optical coherence tomography, thus increasing the
resolution of the system. Later, the work in \cite{gouet2010}
showed that by appropriately introducing a phase conjugator
element in the optical coherence tomography scheme, which produces
a Gaussian-state light source with frequency anti-correlation, a
similar effect could be achieved. In dispersion cancelation, an effect that is observed
in the temporal domain, namely the broadening of the second-order
correlation function of paired photons propagating in two
different optical fibers, it was shown that it can be suppressed, provided that the group velocity
dispersion parameters of both fibers are identical but opposite in
sign, and that the photons are entangled \cite{franson1992,gisin1998,kim2009}. However, it has been recently 
demonstrated that such effects could also be produced by
frequency-correlated photons, which nonetheless might be
non-entangled \cite{torres-company2011, torres-company2012}.

Remote temporal modulation \cite{harris2008,harris2009}, a similar effect to the dispersion
cancelation described above, but observed in the
frequency domain, describes the appearance of new frequency correlations when entangled paired photons are synchronously
driven by two temporal modulators. In a similar manner to dispersion cancelation, if the two
identical modulators are driven in opposite phases, their global
effect is to negate each other, and the spectral correlations
appear as those when no phase modulators are present. Again,
it has been shown \cite{torres-company2012} that entanglement is
not a requisite, and that the same effect can be observed using
non-entangled optical beams bearing certain frequency
correlations.  All these examples illustrate the fact that the presence of
entanglement is not the key enabling factor that allows the
observation of dispersion cancelation and remote temporal
modulation, but the existence of certain frequency correlations, a
characteristic that takes place along the presence of
entanglement, but it can also manifest without it.

In this paper, we consider one important spectroscopic application whose
capabilities have been associated to the use of entangled photon
pairs, namely two-photon virtual-state spectroscopy. The
importance of this technique resides in the fact that, unlike
commonly used two-photon absorption spectroscopy techniques, where
pulsed and tunable sources are required, it is implemented by
carrying out continuous-wave absorption measurements without
changing the wavelength of the source \cite{teich1998,
kojima2004}. Unfortunately, this technique has not been broadly
applied because the ease with which it can be performed is limited
by the low efficiency of spontaneous down conversion in nonlinear
crystals. However, with the advent of ultrahigh flux sources of
entangled photons \cite{dayan_2005}, this technique may open new
research directions towards ultrasensitive detection in chemical and
biological systems \cite{lee_2006,lee_conf}.

The absorption of two photons by an atom or a
molecule induces a transition between two of its energy
levels that match the overall energy of the incident photons. The
quantum mechanical calculation of the TPA transition probability shows
that its value can be understood as a weighted sum of many
energy non-conserving atomic transitions (virtual-state
transitions) \cite{shore1979,sakurai} between energy levels. Then,
the virtual-state transitions, a signature of the medium, can be
revealed experimentally by introducing a delay between the two
absorbed photons, and averaging over different experimental
realizations with different temporal correlations between the
photons \cite{teich1998}. Can we retrieve the sought-after
information (energy level structure) with any type of frequency
correlations between the photons? It has been suggested that
spectroscopic information resident in the TPA signal in
multidimensional pump-probe spectroscopy \cite{mukamel2009} is
essentially the same, regardless of the existence or not of
correlations between the photons absorbed. As stated recently in
\cite{mukamel2012}, {\em it remains however an open question, to
what extent these effects constitute genuine entanglement effects
and whether they can be reproduced, for instance, by shaped or
stochastic classical pulses.}

To unveil the true role of entanglement in virtual-state
spectroscopy, we make use of two ingredients. First, we apply a full
quantum formalism to the two-photon state, so we can identify clearly the amount of
entanglement existing between the photons. Secondly, we consider a
general form of the two-photon state, which allows us to
consider different types of correlations and spectral shapes of the photons.

We will show that the presence of entanglement does not guarantee
the successful retrieval of spectroscopic information of the
medium. In fact, it is the combination of entanglement and a
specific shape of the frequency correlations between photons what
makes the realization of two-photon virtual-state spectroscopy
possible. This result is of great interest because it specifies
the type of two-photon source that needs to be used in order to
experimentally perform the two-photon virtual-state spectroscopy
technique.

\section{Light-matter interaction}
Let us consider the interaction of a medium with a two-photon
optical field $\ket{\Psi}$, described by the interaction
Hamiltonian
$\hat{H}_{I}\pare{t}=\hat{d}\pare{t}\hat{E}^{\pare{+}}\pare{t}$,
where $\hat{d}\pare{t}$ is the dipole-moment operator and
$\hat{E}^{\pare{+}}\pare{t}$ is the positive-frequency part of the
electric-field operator, which reads as
$\hat{E}^{\pare{+}}\pare{t} = \hat{E}^{\pare{+}}_{1}\pare{t} +
\hat{E}^{\pare{+}}_{2}\pare{t}$. The electric field operators
$\hat{E}^{\pare{+}}_{1}\pare{t}$ and
$\hat{E}^{\pare{+}}_{2}\pare{t}$ can be written as
\begin{equation}
\hat{E}^{\pare{+}}_{j}\pare{t} = \int d\omega_{j}
\sqrt{\frac{\hbar \omega_j}{4 \pi \epsilon_0 c
A}}\hat{a}\pare{\omega_{j}}\exp\pare{-i\omega_{j}t},
\label{electric_field1}
\end{equation}
where $c$ is the speed of light, $\epsilon_0$ is the vacuum
permittivity, $A$ is the effective area of the field, and $\hat{a}\pare{\omega_{j}}$ is the annihilation operator
of a photonic frequency-mode with frequency $\omega_{j}$ bearing a specific spatial shape and polarization which, for the sake of simplicity, are not explicitly written.

The medium is initially in its ground state $\ket{g}$ (with
energy $\varepsilon_{g}$). The probability that the medium is
excited to the final state $\ket{f}$ (with energy
$\varepsilon_{f}$), through a two-photon absorption process, is
given by second-order time-dependent perturbation theory as \cite{perina1998}
\begin{equation}\label{probability}
P_{g\rightarrow f} =
\abs{\frac{1}{\hbar^{2}}\int_{-\infty}^{\infty}dt_{2}\int_{-\infty}^{t_{2}}dt_{1}\mbox{M}_{\hat{d}}\pare{t_{1},t_{2}}\mbox{M}_{\hat{E}}\pare{t_{1},t_{2}}}^{2},
\end{equation}
with
\begin{eqnarray}
\label{M_d_1}
\mbox{M}_{\hat{d}}\pare{t_{1},t_{2}} &=& \bra{f}\hat{d}\pare{t_{2}}\hat{d}\pare{t_{1}}\ket{g}, \\
\label{M_E_1} \mbox{M}_{\hat{E}}\pare{t_{1},t_{2}} &=&
\bra{\Psi_f}\hat{E}^{\pare{+}}\pare{t_{2}}\hat{E}^{\pare{+}}\pare{t_{1}}\ket{\Psi},
\end{eqnarray}
where $\ket{\Psi_{f}}$ denotes the final state of the optical
field.

Equation (\ref{M_d_1}) can be expanded in terms of virtual-state
transitions, to obtain
\begin{eqnarray}
\mbox{M}_{\hat{d}}\pare{t_{1},t_{2}} =& \sum_{j=1}&D^{\pare{j}}\exp\cor{{-i\pare{\varepsilon_{j}-i\kappa_{j}/2-\varepsilon_{f}}t_{2}}} \nonumber \\
&\times& \exp\cor{-i\pare{\varepsilon_{g}-\varepsilon_{j}+i\kappa_{j}/2}t_{1}},
\label{M_d_2}
\end{eqnarray}
where $D^{\pare{j}}=\bra{f}\hat{d}\ket{j}\bra{j}\hat{d}\ket{g}$
are the transition matrix elements of the dipole-moment operator.
Equation (\ref{M_d_2}) shows that the excitation of the medium occurs
through intermediate states $\ket{j}$, with complex energy
eigenvalues $\varepsilon_{j}-i\kappa_{j}/2$, where $\kappa_{j}$
takes into account the natural linewidth of the intermediate
states \cite{mollow1968}. Also, we can write Eq. (\ref{M_E_1}) as
\begin{eqnarray}
\mbox{M}_{\hat{E}}\pare{t_{1},t_{2}} = \bra{\psi_f}\hat{E}_{2}^{\pare{+}}\pare{t_{2}}\hat{E}_{1}^{\pare{+}}\pare{t_{1}}\ket{\psi_{i}}+ \bra{\psi_f}\hat{E}_{1}^{\pare{+}}\pare{t_{2}}\hat{E}_{2}^{\pare{+}}\pare{t_{1}}\ket{\psi_{i}},
\label{M_E_2}
\end{eqnarray}
where we have kept the terms in which only one photon from each
field contributes to the overall two-photon excitation. The first
term of Eq. (\ref{M_E_2}) corresponds to the case in which the
photon field $\hat{E}_{1}^{\pare{+}}\pare{t}$ interacts first, and
$\hat{E}_{2}^{\pare{+}}\pare{t}$ interacts later. The remaining
term describes the complementary case.

Since we are interested in a two-photon absorption process, we
consider the initial state of the optical field as an arbitrary
two-photon state, which can be written as \cite{juan2011}
\begin{equation}
\label{mode_func0} \ket{\Psi}=\int
d\nu_{s}d\nu_{i}\Phi\pare{\nu_{s},\nu_{i}}\hat{a}_s^{\dagger}\pare{\nu_{s}+\omega_{s}^{0}}\hat{a}_i^{\dagger}\pare{\nu_{i}+\omega_{i}^{0}}\ket{0},
\end{equation}
where $s$ and $i$ stand for signal and idler photonic modes, $\nu_{j}=\omega_{j}-\omega_{j}^{0}$ ($j=s,i$) are the frequency
deviations from the central frequencies $\omega_{j}^{0}$, and
$\Phi\pare{\nu_{s},\nu_{i}}$ is the joint spectral
amplitude, or mode function, which fully describes the correlations
and bandwidth of the two-photon state.

Finally, to quantify the degree of entanglement between the absorbed photons, we make use of
the entropy of entanglement, defined as \cite{law2000}
\begin{equation}
E=-\sum_{i}\lambda_{i}\mbox{log}_{2}\lambda_{i},
\end{equation}
where $\lambda_{i}$ are the eigenvalues of the Schmidt decomposition of the joint spectral amplitude, i.e., $\Phi\pare{\nu_{s},\nu_{i}}=\sum_i
\sqrt{\lambda_i} f_i(\nu_s) g(\nu_i)$, with $f$ and $g$ corresponding to the
Schmidt modes. It is worth remarking that the lack of entanglement between the pair of photons is characterized by a value of the entropy equal to zero.

\section{Two-photon absorption transition probability}
With the aim of recognizing in which situations virtual-state
spectroscopy can be performed, we will compute the TPA transition
probability of atomic hydrogen using different types of initial
two-photon states. We have selected atomic hydrogen as model
system, because it has been used in previous studies of
virtual-state spectroscopy \cite{perina1998, teich1998} and it has
been the subject of several one- and two-photon absorption
experiments \cite{etherton1970, cesar1996, book_bethe, bebb1966}.
In our calculations, we will focus on the $1s\rightarrow2s$
two-photon transition. Due to quantum number selection rules
\cite{book_bethe}, this transition takes place via intermediate
$p$ states: $1s\rightarrow\llav{2p, 3p, ..., np}\rightarrow2s$,
which are coupled to the $s$ states by real-valued transition
matrix elements. The hydrogen atom energy levels are
$\varepsilon_{n}=-13.6/n^{2}$ eV ($n=1,2,3,...$) and the natural
linewidths of intermediate states $\kappa_{j}$ are taken from
Refs. \cite{etherton1970,book_bethe}. We assume the condition
$\varepsilon_{f}-\varepsilon_{g} = \omega_{s}^{0} +
\omega_{i}^{0}$, and that the final state $2s$ is Lorentzian
broadened with a radiative lifetime $1/\kappa_{f} = 122$ ms
\cite{cesar1996}, which is introduced in the model by averaging
the TPA transition probability over a Lorentzian function of width
$\kappa_{f}$ \cite{mollow1968}.

\subsection{TPA transition probability with uncorrelated classical pulses}

Let us consider first the case that has been studied in Ref.
\cite{mukamel2009}. It corresponds to the situation in which the
two absorbed photons are embedded into rectangular-shaped pulses
of the same duration $T_{p}$, with a tunable time delay $\tau$
between them. This initial optical field can be represented by an
uncorrelated two-photon state described by the normalized mode
function
\begin{equation} \label{mode_func_unc}
\Phi\pare{\nu_{s},\nu_{i}} = \frac{T_{p}}{2\pi}\mbox{sinc}\pare{T_{p} \nu_{s}/2} \mbox{sinc}\pare{T_{p}\nu_{i}/2} \exp \cor{ i \pare{\nu_s-\nu_i} \tau /2}.
\end{equation}
With the state given by Eq. (\ref{mode_func_unc}), and making use
of Eqs. (\ref{probability}), (\ref{M_d_2}) and (\ref{M_E_2}), we can
write the TPA transition probability as
\begin{equation} \label{pro_unc}
P_{g \rightarrow f}\pare{T_{p};\tau} = \frac{\omega_{0}^{2}}{\hbar^{2}\epsilon_{0}^{2}c^{2}A^{2}T_{p}^{2}}\abs{\sum_{j}D^{\pare{j}}\cor{I_{1}+I_{2}}}^{2},
\end{equation}
where
\begin{eqnarray}
I_{1} = &\frac{\sin\cor{\Delta\omega\pare{T_{p}-\tau}/2}}{\Delta_{g} \Delta\omega} - \frac{\sin\cor{\Delta_{f}\pare{T_{p}-\tau}/2}\exp\cor{{i\Delta_{g}\pare{T_{p}+\tau}/2}}}{\Delta_{g} \Delta_{f}} \nonumber \\
&- \frac{2i\sin\pare{\Delta_{g}T_{p}/2}\sin\pare{\Delta_{f}\tau /2}\exp\cor{-i\pare{ \Delta_{f}T_{p} - \Delta_{g}\tau}/2 }} {\Delta_{g} \Delta_{f}} ,
\end{eqnarray}
\begin{eqnarray}
I_{2} = \frac{\sin\cor{\Delta\omega\pare{T_{p}-\tau}/2}}{\Delta_{g} \Delta\omega} - \frac{\sin\cor{\Delta_{f}\pare{T_{p}-\tau}/2}\exp\cor{{i\Delta_{g}\pare{T_{p}-\tau}/2}}}{\Delta_{g} \Delta_{f}} ,
\end{eqnarray}
with $\Delta_{f} =
\varepsilon_{j}-i\kappa_{j}/2-\varepsilon_{f}+\omega_{0}$,
$\Delta_{g} =
\varepsilon_{g}-\varepsilon_{j}+i\kappa_{j}/2+\omega_{0}$ and
$\Delta\omega = \varepsilon_{g}-\varepsilon_{f}+2\omega_{0}$. For
the sake of simplicity, we have assumed the condition
$\omega_{i}^{0}=\omega_{s}^{0}=\omega_{0}$.

We have computed the TPA transition probability for different
values of $T_p$ and $\tau$. In all cases, it turns out be constant
as a function of the delay ($\tau$) between the pulses, when $\tau
< T_p$, which implies that a Fourier analysis
with respect to $\tau$ would result in only one peak centered at
zero-frequency, meaning that spectroscopic information about
intermediate levels of the medium is not present in the TPA
signal.

From these results one can infer that when frequency correlations
between the photons are not present, spectroscopic information
about energy levels is not available. This implies that
virtual-state spectroscopy cannot be performed by means of two
delayed rectangular-shaped classical pulses, which is in contradiction with the
results presented in section VI of ref. \cite{mukamel2009, contradiction}.

\subsection{TPA transition probability with classically frequency-correlated photons}

In this section, we explore the case in which the two absorbed
photons are frequency-correlated but they are nonetheless non-entangled. To
this end, we make use of the theory presented by Mollow in
\cite{mollow1968} and rewrite the TPA transition probability [Eq.
(\ref{probability})] as
\begin{equation}
P_{g\rightarrow f} = \frac{1}{\hbar^{4}}\int_{-\infty}^{\infty} dt_{2}^{'}dt_{1}^{'}dt_{2}dt_{1}\mathcal{L}^{*}(t_{2}^{'},t_{1}^{'})G^{(2)}(t_{2}^{'},t_{1}^{'};t_{2},t_{1})\mathcal{L}(t_{2},t_{1}),
\end{equation}
where $\mathcal{L}(t_{2},t_{1}) = \Theta\pare{t_{2}-t_{1}}\mbox{M}_{\hat{d}}\pare{t_{1},t_{2}}$, with $\Theta\pare{t}$ being the Heaviside step function. Here, $G^{\pare{2}}$ corresponds to the second-order field correlation function, which is defined in terms of the density operator $\hat{\rho}$ of the optical field as
\begin{eqnarray}\label{G2}
G^{(2)}(t_{2}^{'},t_{1}^{'};t_{2},t_{1}) =& \mbox{Tr}\cor{\hat{\rho}\hat{E}^{\pare{-}}_{2}(t_{2}^{'})\hat{E}^{\pare{-}}_{1}(t_{1}^{'})\hat{E}^{\pare{+}}_{1}\pare{t_{2}}\hat{E}^{\pare{+}}_{2}\pare{t_{1}}} \nonumber \\
&+\mbox{Tr}\cor{\hat{\rho}\hat{E}^{\pare{-}}_{2}(t_{2}^{'})\hat{E}^{\pare{-}}_{1}(t_{1}^{'})\hat{E}^{\pare{+}}_{2}\pare{t_{2}}\hat{E}^{\pare{+}}_{1}\pare{t_{1}}} \nonumber \\
&+\mbox{Tr}\cor{\hat{\rho}\hat{E}^{\pare{-}}_{1}(t_{2}^{'})\hat{E}^{\pare{-}}_{2}(t_{1}^{'})\hat{E}^{\pare{+}}_{1}\pare{t_{2}}\hat{E}^{\pare{+}}_{2}\pare{t_{1}}} \nonumber \\
&+\mbox{Tr}\cor{\hat{\rho}\hat{E}^{\pare{-}}_{1}(t_{2}^{'})\hat{E}^{\pare{-}}_{2}(t_{1}^{'})\hat{E}^{\pare{+}}_{2}\pare{t_{2}}\hat{E}^{\pare{+}}_{1}\pare{t_{1}}},
\end{eqnarray}
where $\mbox{Tr}\cor{...}$ stands for the trace over the field states.

To compute the second-order correlation function, we consider a
classically-correlated two-photon state described by a density
operator of the form
\begin{equation}\label{density}
\hat{\rho} = \int d\nu \abs{\Phi\pare{\nu,-\nu}}^{2}\vert\omega^{0}+\nu \rangle_{1}\vert\omega^{0}-\nu\rangle_{2}\langle\omega^{0}+\nu\vert_{1}\langle\omega^{0}-\nu\vert_{2},
\end{equation}
with $\omega^{0}$ being the central frequency of the photons and
$\Phi\pare{\nu,-\nu}$ the mode function that describes the
frequency correlations between them.

By using Eq. (\ref{density}) we find that the second-order
correlation function of the classically correlated photons is
given by

\begin{eqnarray}
G^{2}(t_{2}^{'},t_{1}^{'};t_{2},t_{1}) = &\pare{\frac{\hbar\omega ^{0}}{2\pi\epsilon_{0}c A}}^{2}\exp[i\omega^{0}(t_{2}^{'}+t_{1}^{'}-t_{2}-t_{1})] \nonumber \\
&\times \int d \nu \abs{\Phi\pare{\nu,-\nu}}^2\cos\cor{\nu\pare{t_{2}-t_{1}}}\cos [\nu(t_{2}^{'}-t_{1}^{'})].
\end{eqnarray}

Notice that the presence of the norm of the mode function cancels
out the phase difference introduced by the delay $\tau$ [see Eqs.
(\ref{mode_func_unc}), (\ref{gauss_mod}) and (\ref{sinc_mod})].
Consequently, the TPA transition probability does not depend on
the delay between the photons, which implies that when using
non-entangled frequency-correlated photons, spectroscopic
information about intermediate levels of the medium is not
available in the TPA signal.

\subsection{TPA transition probability with entangled photons}

In view of the previous results, and the ideas and
calculations presented originally in \cite{teich1998}, it naturally arises the
question if the presence of a high-degree of frequency
entanglement between the photons is the key ingredient that allows
to access information about the energy level structure of a
medium by means of two-photon virtual-state spectroscopy. In what
follows, we will show that the use of highly entangled
photons does not guarantee the successful retrieval of
spectroscopic information of the medium. Rather, the use of a specific spectral shape of the frequency correlations is what
makes the realization of two-photon virtual-state spectroscopy
possible, even when quasi-uncorrelated paired photons (low degree
of entanglement) are considered.

\subsubsection{Two-photon state with a Gaussian spectral shape}

In general, a two-photon state with tunable frequency
correlations, and consequently tunable degree of entanglement, can
be generated by means of type-II Spontaneous Parametric Down
Conversion (SPDC), where two photons with orthogonal polarizations
are generated in a second-order nonlinear crystal of length $L$,
when pumped by a Gaussian pulse with temporal duration $T_{+}$. After the
crystal, signal and idler photons interchange their polarization
and traverse a similar crystal of length $L/2$. After the addition
of a tunable delay $\tau$ between the photons, and restricting
their spectrum using a Gaussian filter, the normalized mode
function reads as
\begin{eqnarray}\label{gauss_mod}
\Phi\pare{\nu_{s},\nu_{i}} = & \pare{\frac{T_{-}T_{+}}{\sqrt{2\pi}}}^{1/2} \exp\cor{-T_{+}^{2}\pare{\nu_{i}+\nu_{s}}^{2}} \exp\cor{-T_{-}^{2}\pare{\nu_{s}-\nu_{i}}^{2}/4} \nonumber \\
& \times \exp\cor{iLN_{p}\pare{\nu_{s}+\nu_{i}}/2 + i\nu_{i}\tau},
\end{eqnarray}
where $T_{-}=\pare{N_{s}-N_{i}} L/2$, $N_{j}$ ($j=i,s,p$) are the
inverse group velocities. We have made use of the group velocity
matching condition $N_{p}=\pare{N_{i}+N_{s}}/2$ \cite{keller1997},
which ease the tuning of the frequency correlations, and the
degree of entanglement, between the photons \cite{hendrych2007}.

\begin{figure}[!t]\label{spectrum}
    \begin{center}
       \subfigure[]{\includegraphics[width=4.7cm]{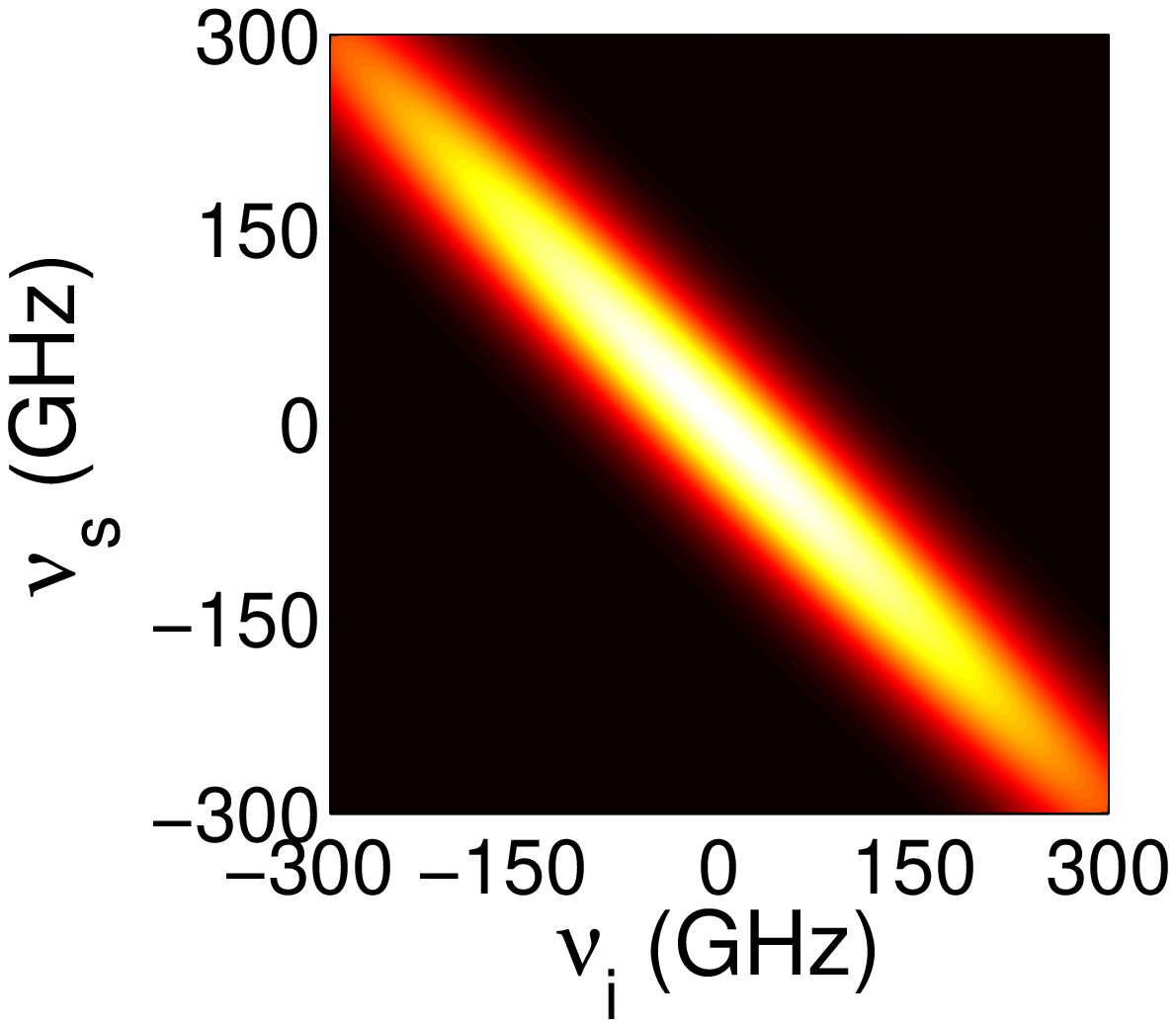}} \hspace{-5mm}
       \subfigure[]{\includegraphics[width=4.7cm]{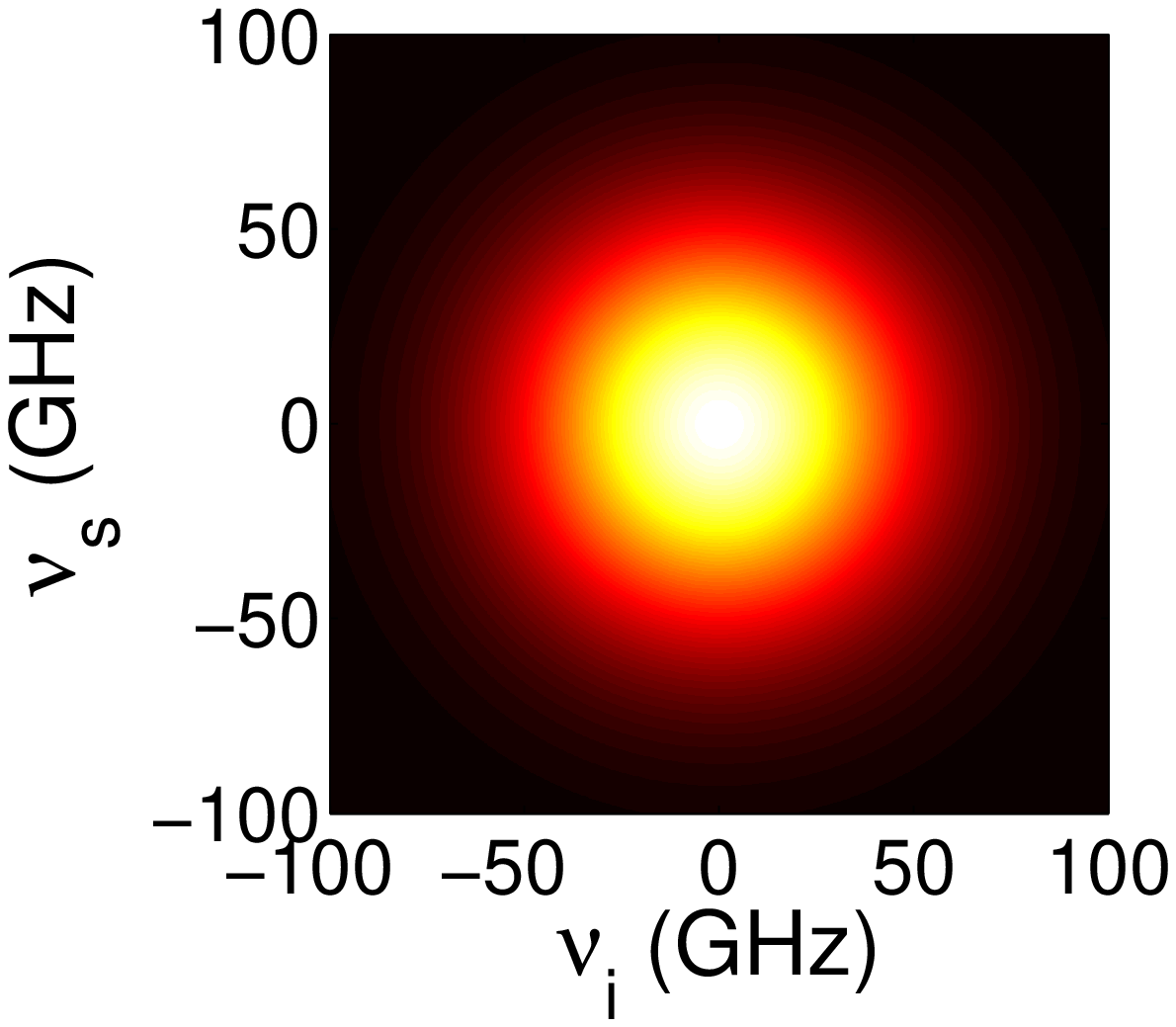}} \hspace{-5mm}
       \subfigure[]{\includegraphics[width=4.7cm]{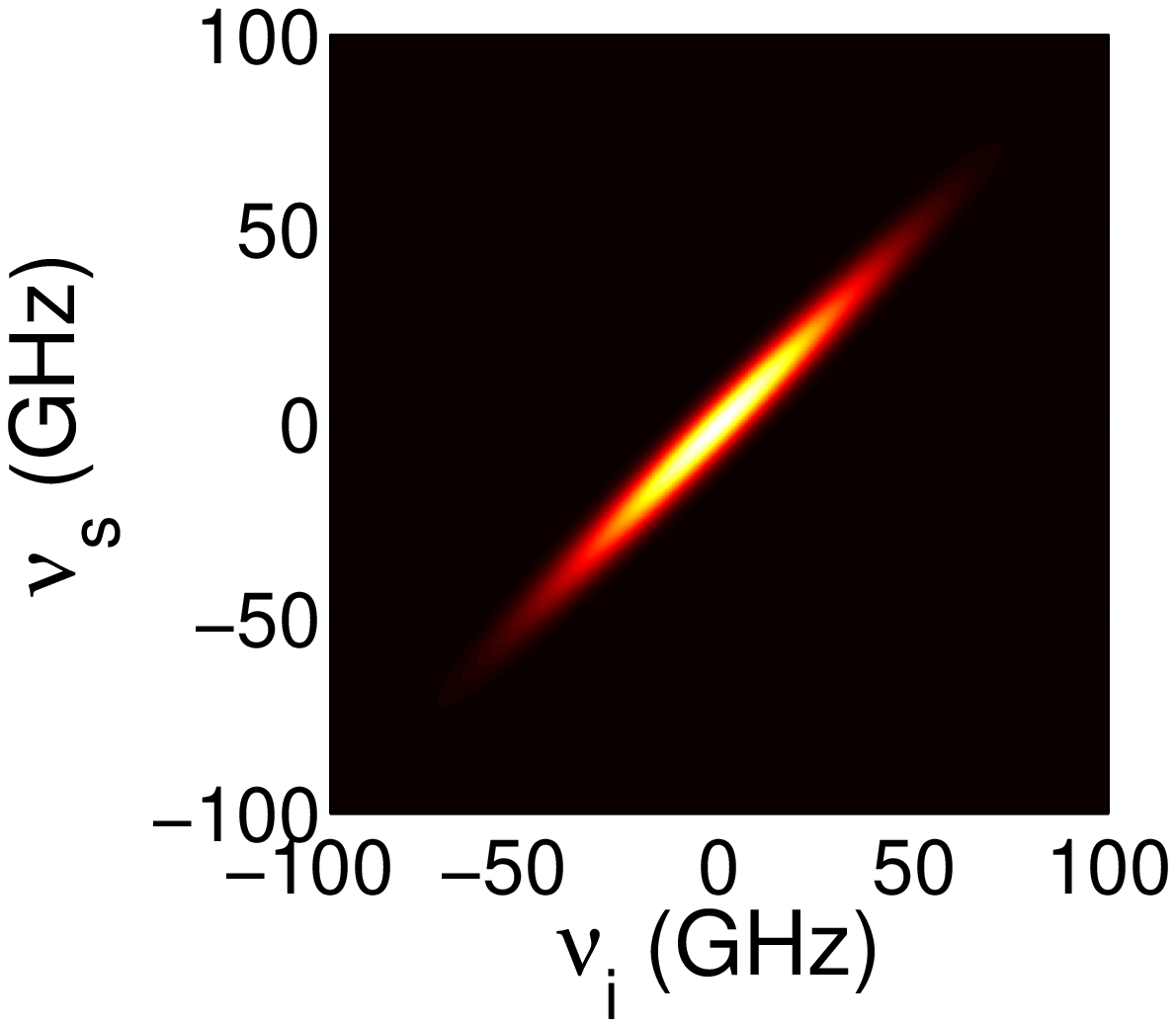}} \\
       \subfigure[]{\includegraphics[width=8cm]{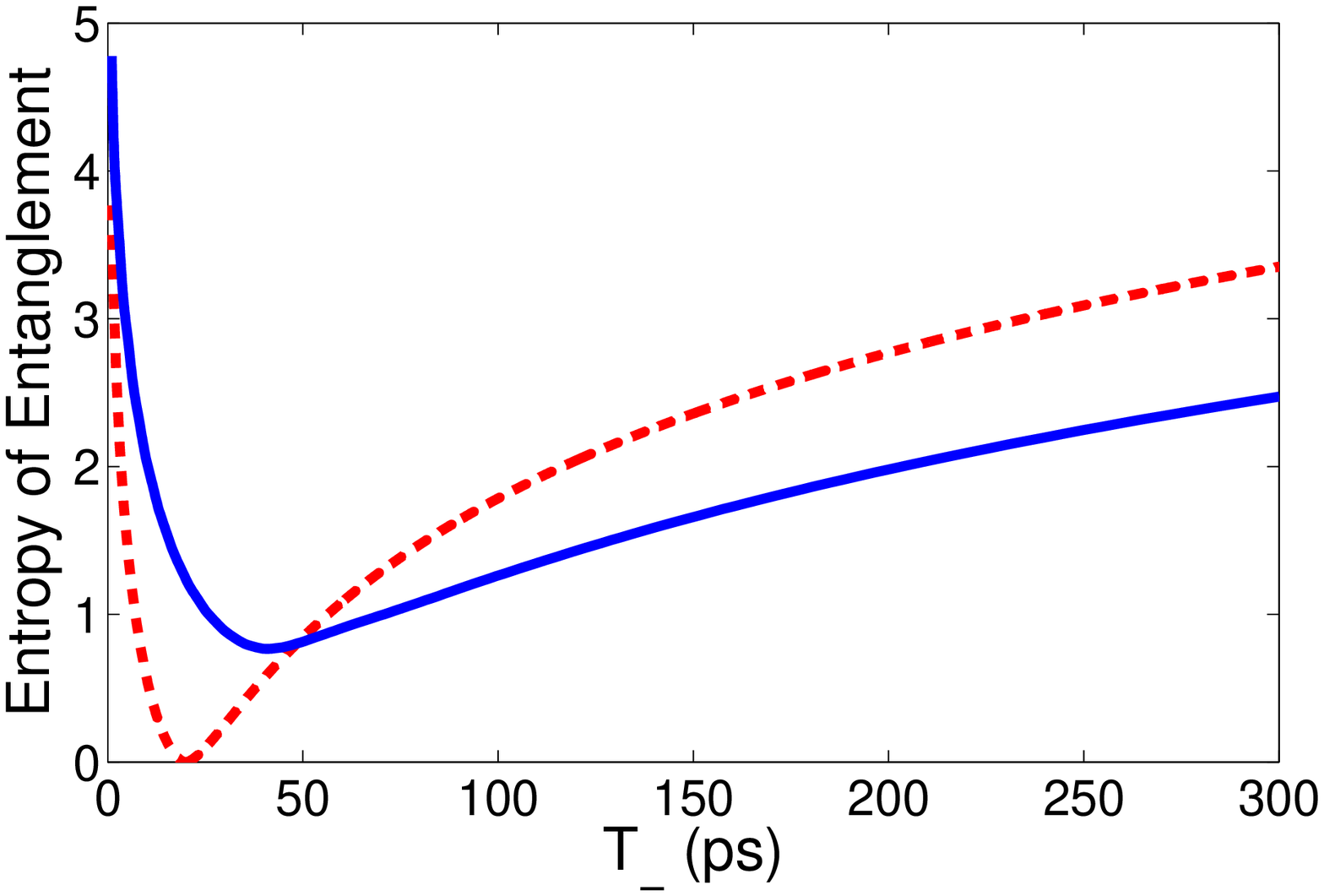}}
    \end{center}
\caption{Joint spectrum of the two-photon state for different
values of $T_{-}$: (a) $T_{-}=2\;\mbox{ps}$, (b)
$T_{-}=20\;\mbox{ps}$, and (c) $T_{-}=200\;\mbox{ps}$. (d) Entropy
of entanglement as a function of $T_{-}$ for Gaussian (red dashed
line) and sine cardinal (blue solid line) shapes of the mode
function. In all cases, the pump pulse duration is: $T_{+}=10 \;\mbox{ps}$. }
\end{figure}

The frequency correlations of the down-converted photons can be
tuned by carefully selecting the values of $T_{+}$ and $T_{-}$.
Figures 1(a) to 1(c) show the joint probability distribution of
the two-photon state $S\pare{\nu_{s},\nu_{i}} =
\abs{\Phi\pare{\nu_{s},\nu_{i}} }^{2}$, which measures the
probability of detecting a signal photon of frequency
$\omega_{s}^{0}+\nu_{s}$ in coincidence with an idler photon of
frequency $\omega_{i}^{0}+\nu_{i}$. Frequency anti-correlated
photons [Fig. 1(a)] ($\nu_s \sim -\nu_i$) are obtained when
$T_{+}\gg T_{-}$, whereas for $T_{+}\ll T_{-}$, we obtain
frequency correlated photons [Fig. 1(c)]. In the particular case
when $T_{-}=2T_{+}$, frequency uncorrelated pairs of photons [Fig.
1(b)] are generated. Figure 1(d) (red dashed line) shows the
dependence of the entropy of entanglement with $T_{-}$ for a fixed
value of $T_{+}$, for a mode function of the form given by Eq.
(\ref{gauss_mod}).

Using the initial two-photon state described by Eq.
(\ref{gauss_mod}), we find that the TPA transition probability is
given by
\begin{eqnarray}\label{prob_gauss}
P_{g\rightarrow f}\pare{T_{-},T_{+};\tau} =& \frac{32\pi\omega_{0}^{2}}{\hbar^{2}\epsilon_{0}^{2}c^{2}A^{2}}T_{+}T_{-}\exp\cor{-2T_{+}^{2}\pare{\varepsilon_{g}-\varepsilon_{f}+\omega_{p}}^{2}} \nonumber \\
&\times \left\vert \sum_{j}D^{\pare{j}}\left\{F_{+}\cor{\eta^{\pare{j}}T_{-};\tau}\exp\cor{-i\eta^{\pare{j}}\tau}\right.\right. \nonumber \\
& \left. \textcolor{white}{\sum_{j}} \left. + F_{-}\cor{\eta^{\pare{j}}T_{-};\tau}\exp\cor{i\eta^{\pare{j}}\tau}\right\}\right| ^{2},
\end{eqnarray}
where $\eta^{\pare{j}} = \Delta^{\pare{j}} - i\kappa_{j}/2$, with the energy mismatch given by $\Delta^{\pare{j}} = \varepsilon_{j}-\varepsilon_{g}-\omega_{0}$, and the function $F$ defined as
\begin{equation}
F_{\pm}\pare{\xi;\tau} = \exp\pare{-\xi^{2}}\cor{1-\frac{2i}{\sqrt{\pi}}\int_{0}^{\xi\pm\frac{i\tau}{2T_{-}}}\exp\pare{y^{2}}dy}.
\end{equation}

We have computed the TPA transition probability as a function of the delay
between photons considering states bearing different types of
correlations, particularly for uncorrelated and anti-correlated pairs of photons. As previously obtained, in the
case of uncorrelated photons, the TPA signal is
constant with the delay $\tau$, so no spectroscopic information
about intermediate levels is available.

Surprisingly, in the case of anti-correlated photons, the TPA
transition probability is also constant with the delay $\tau$,
which means that information about the energy level structure of
the medium cannot be retrieved from the TPA signal either. This
result is of great interest since it tells us that the use of a
source of paired photons with entanglement does not guarantee the
successful retrieval of such information. We need to consider
another property of the two-photon state that is needed in order
to perform virtual-state spectroscopy, namely a specific spectral
shape of the frequency correlations.

\subsubsection{Two-photon state with a Sine cardinal spectral shape}

Fortunately, two-photon states with a Gaussian shape, which
require a strong filtering of the pair of photons
\cite{nakanishi2009}, are not naturally harvested in SPDC. By
considering a more realistic shape of the mode function, we will
show that two-photon virtual-state spectroscopy can retrieve the
sought-after information about the energy level structure under a
great variety of circumstances.

As in the previous section, we consider a
type-II SPDC process where an additional nonlinear crystal of length $L/2$ is used to
achieve group velocity compensation. By introducing a tunable
delay $\tau$ between the photons, without restricting their
spectrum, the normalized mode function is written as
\begin{eqnarray}\label{sinc_mod}
\Phi\pare{\nu_{s},\nu_{i}} =& \pare{\frac{T_{-}T_{+}}{2\pi\sqrt{2\pi}}}^{1/2}\exp\cor{-T_{+}^{2}\pare{\nu_{i}+\nu_{s}}^{2}} \mbox{sinc}\cor{T_{-}\pare{\nu_{s}-\nu_{i}}/2} \nonumber \\
& \times \exp\cor{iLN_{p}\pare{\nu_{s}+\nu_{i}}/2 + i\nu_{i}\tau}.
\end{eqnarray}
The entropy of entanglement of the two-photon state described by Eq.
(\ref{sinc_mod}) is shown in Fig. 1(d) (blue solid line). Notice
that in this case, due to the presence of the sine cardinal
function, only quasi-uncorrelated photons can be generated.

We now make use of the initial two-photon state described by the
mode function given in Eq. (\ref{sinc_mod}) to write the TPA
transition probability as
\begin{eqnarray}\label{prob_sinc}
P_{g\rightarrow f}\pare{T_{-},T_{+};\tau} = & \frac{64\pi\omega_{0}^{2}}{\hbar^{2}\epsilon_{0}^{2}c^{2}A^{2}T_{-}}\cor{\frac{\sqrt{2}T_{+}}{\sqrt{\pi}}\exp\cor{-2T_{+}^{2}\pare{\varepsilon_{g}-\varepsilon_{f}+\omega_{p}}^{2}}} \nonumber \\
& \times \left\vert \sum_{j}A^{\pare{j}}\left\{ 2 - \exp\cor{-i\eta^{\pare{j}}\pare{T_{-}-\tau}}\right.\right. \nonumber \\
& \left. \textcolor{white}{\sum_{j}} \left. - \exp\cor{-i\eta^{\pare{j}}\pare{T_{-}+\tau}}\right\} \right\vert ^{2},
\end{eqnarray}
where $A^{\pare{j}} = D^{\pare{j}}/\eta^{\pare{j}}$.

\begin{figure}[t!]
    \begin{center}
       \includegraphics[width=8.2cm]{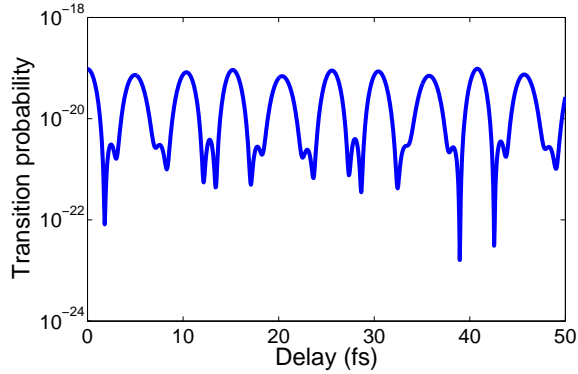}
    \end{center}
\caption{Transition probability as a function of the delay $\tau$
for anti-correlated photons ($T_{-}=2$ ps). Pump pulse duration:
$T_{+}=10$ ps. Y-axis in logarithmic scale.}
\end{figure}

Figure 2 shows the TPA transition probability as a function of the
delay between the pulses. Notice the nonmonotonic behavior of the
TPA transition probability for anti-correlated photons. This means
that spectroscopic information is contained within the TPA signal,
which might be related to the energy level structure of the
medium. In order to retrieve this information, we follow
\cite{teich1998} and perform an average of Eq. (\ref{prob_sinc})
over a range of values of $T_{-}$ to obtain the
weighted-and-averaged TPA transition probability
\begin{equation}\label{weighted}
\bar{P}\pare{\tau} = \frac{1}{T}\int_{T_{-}^{\mbox{\scriptsize{min}}}}^{T_{-}^{\mbox{\scriptsize{max}}}} P_{g\rightarrow f}\pare{T_{-},T_{+};\tau}T_{-}dT_{-},
\end{equation}
where $T=T_{-}^{\mbox{\scriptsize{max}}} - T_{-}^{\mbox{\scriptsize{min}}}$.

To experimentally perform the average in Eq. (\ref{weighted}), a
set of experiments with different values of $T_{-}$ are needed.
Fortunately, parameter $T_{-}$ can be tuned over a relatively
broad range by using different methods, depending on the system
configuration. For instance, in type-I SPDC (parallel-polarized
photons), changing the width of the pump beam modifies the value
of $T_{-}$ \cite{joobeur1994}, whereas in type-II, $T_{-}$ is
linearly proportional to the crystal length \cite{shih}, so a
proper set of wedge-shaped nonlinear crystals might be used.

\begin{figure}[t!]\label{spectroscopy}
    \begin{center}
       \includegraphics[width=8.3cm]{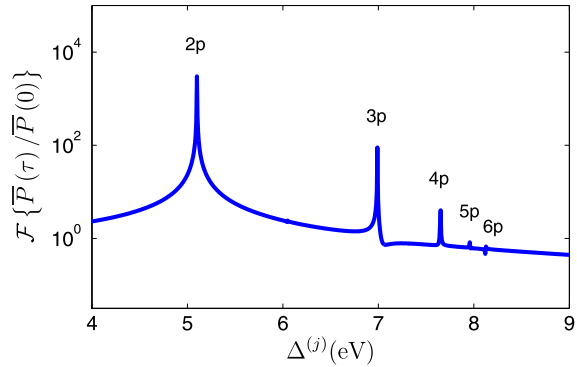}
    \end{center}
\caption{Fourier transform of the normalized weighted-and-averaged
TPA transition probability as a function of the energy mismatch
$\Delta^{\pare{j}}$. The delay range considered is $0\leq\tau\leq
2\;\mbox{ps}$, with an integration time of $2\leq T_{-}\leq 10$
ps. Y-axis in logarithmic scale.}
\end{figure}

\begin{figure}[t!]\label{spectrum}
    \begin{center}
       \subfigure[]{\raisebox{-39.5mm}{\includegraphics[width=5.5cm]{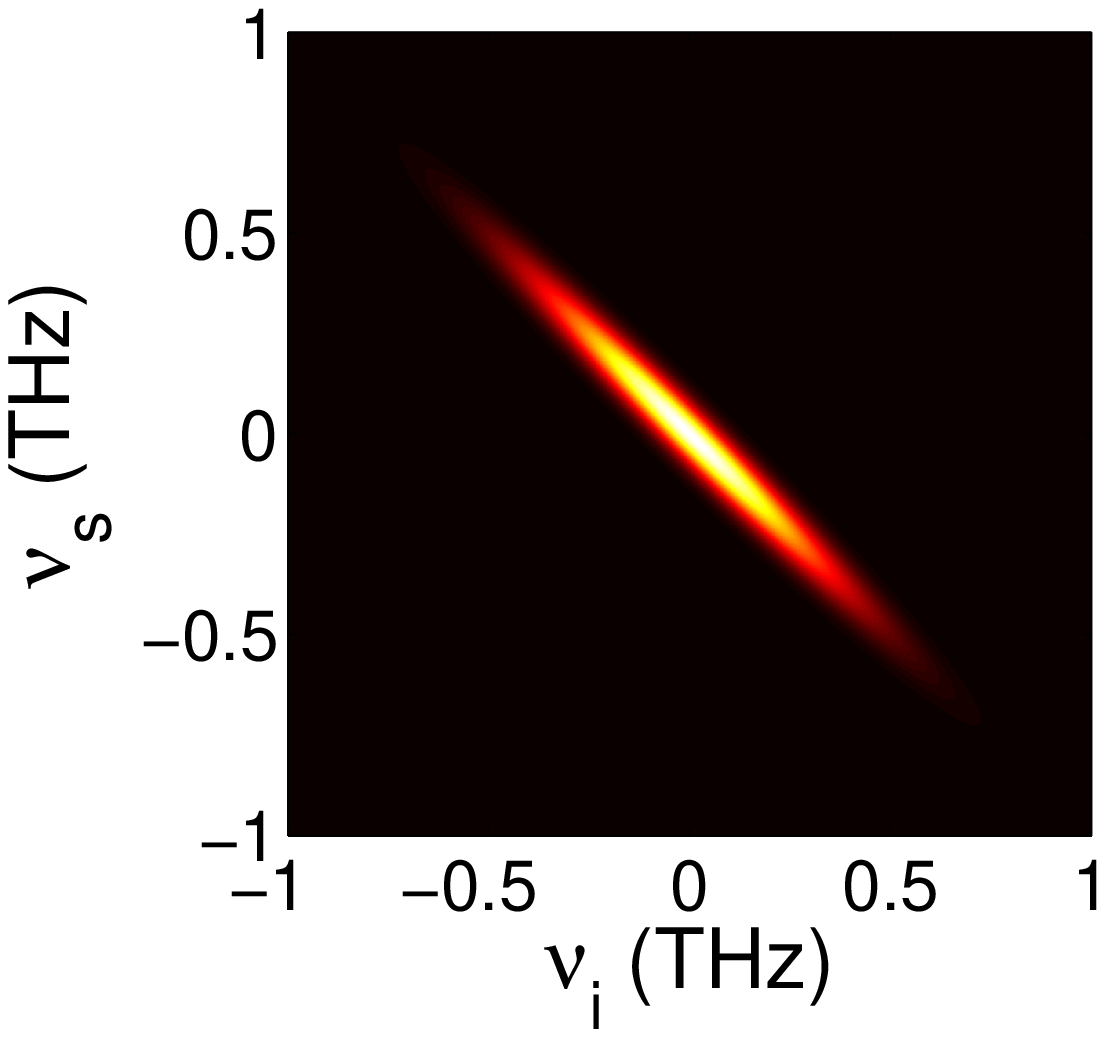}}}
       \subfigure[]{\includegraphics[width=7.5cm]{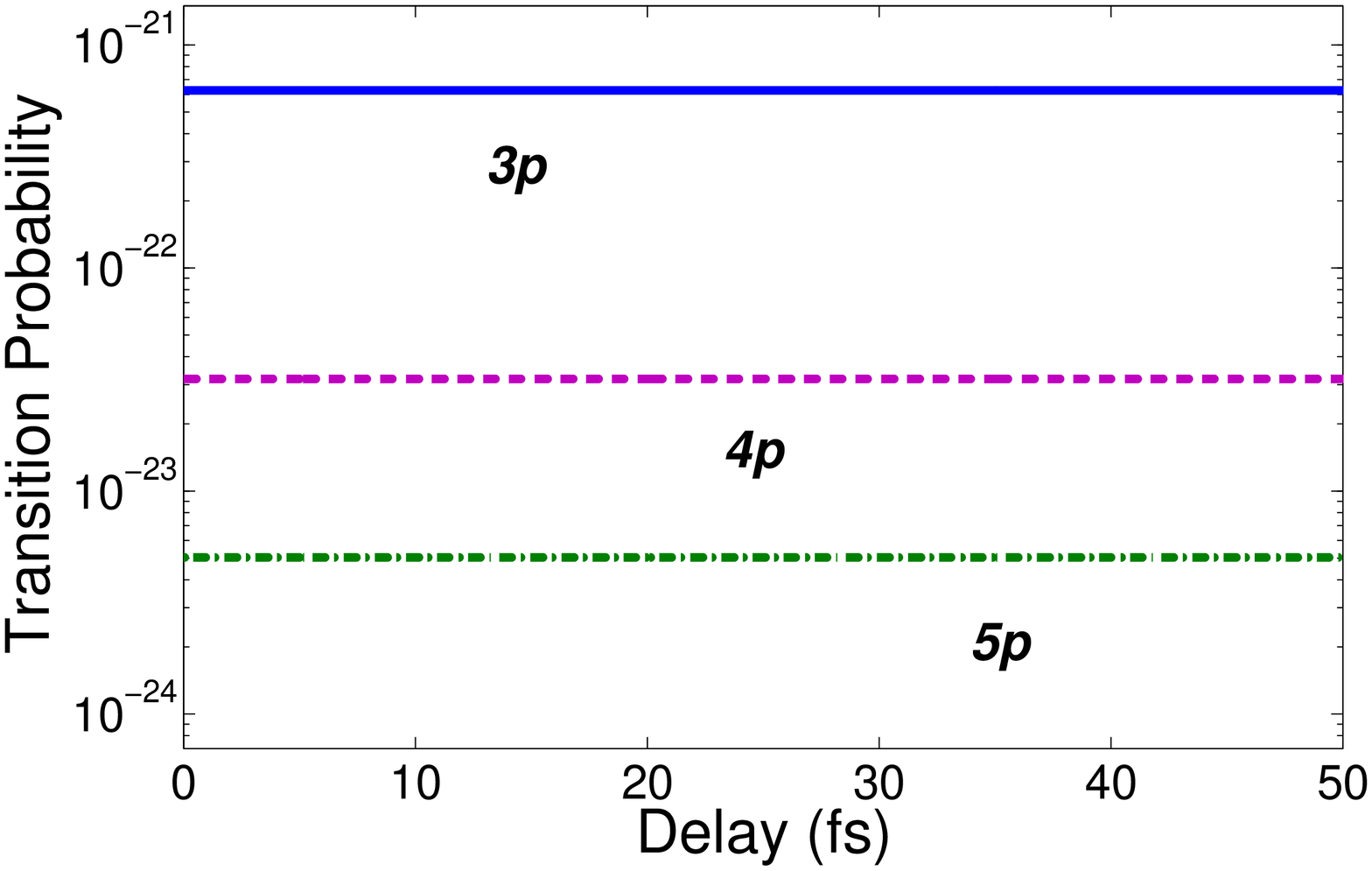}}\\
       \subfigure[]{\raisebox{-39.5mm}{\includegraphics[width=5.5cm]{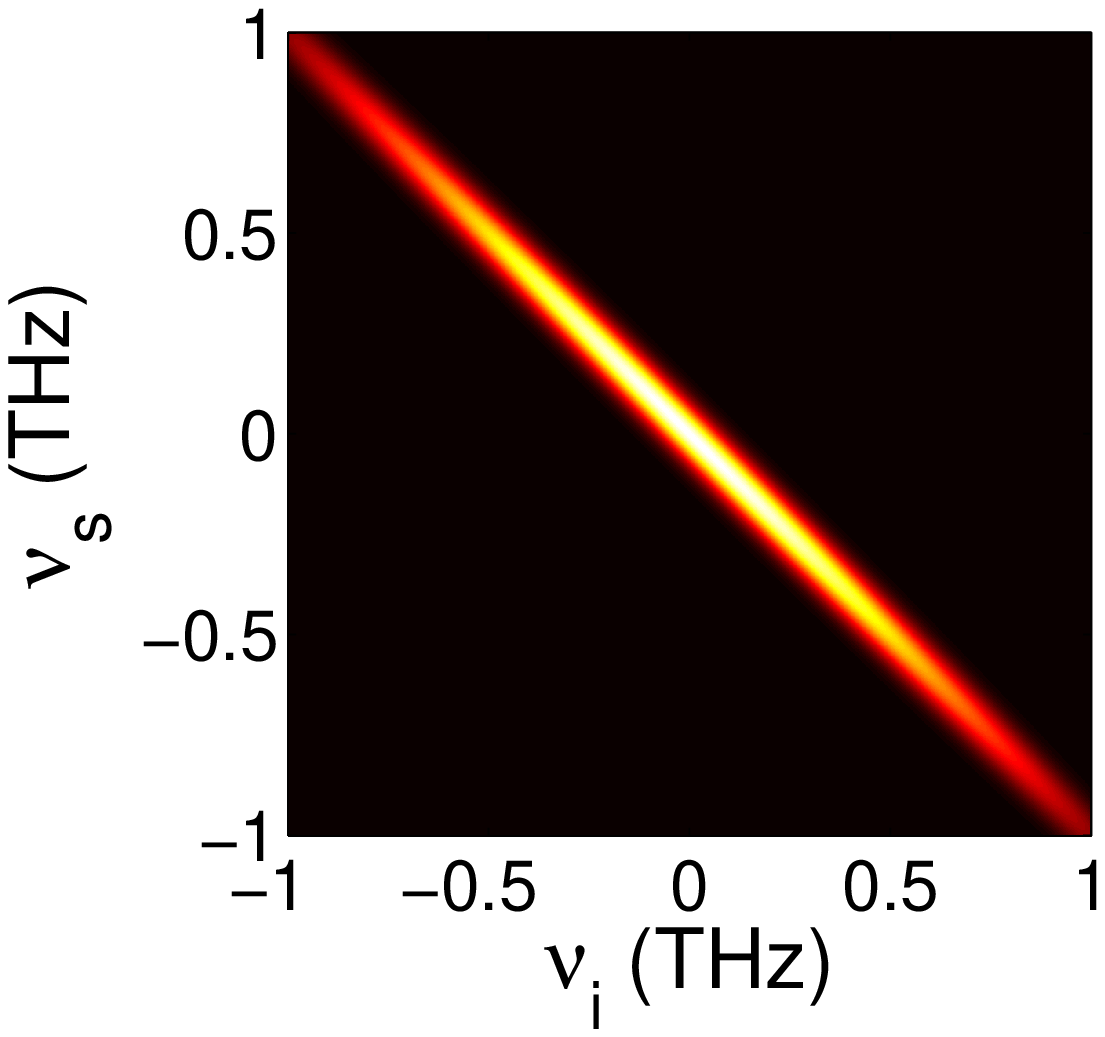}}}
       \subfigure[]{\includegraphics[width=7.5cm]{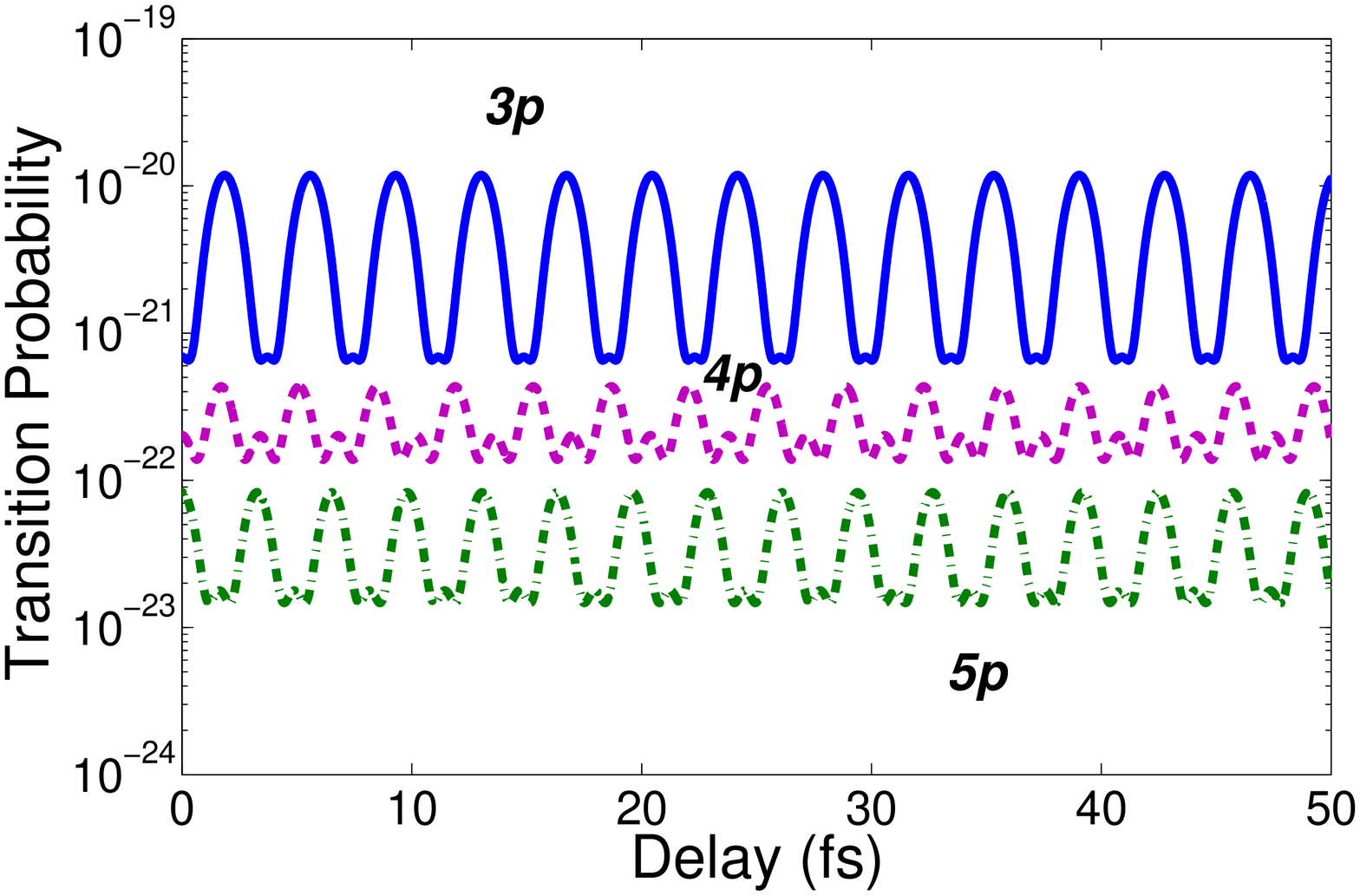}}\\
       \subfigure[]{\raisebox{-39.5mm}{\includegraphics[width=5.5cm]{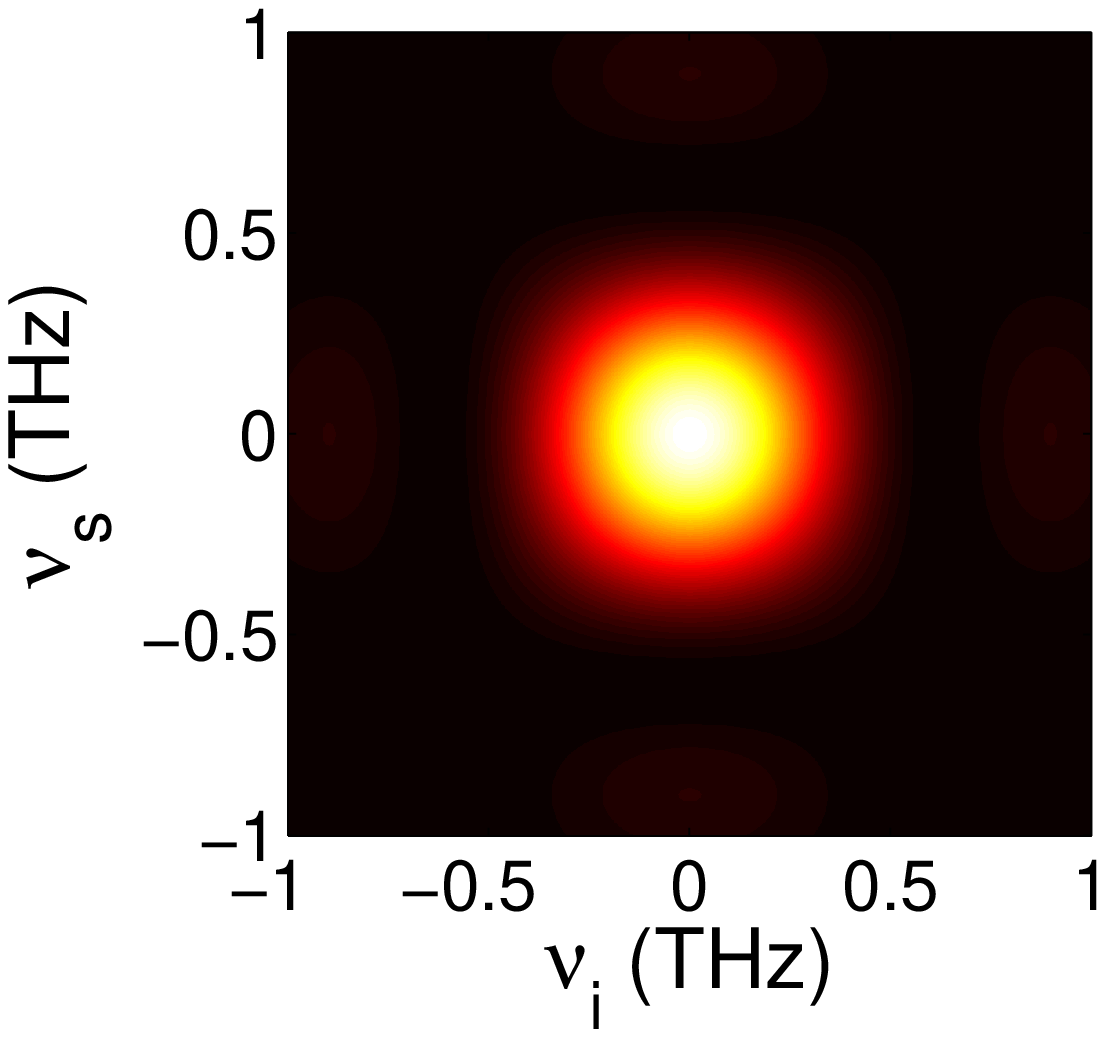}}}
       \subfigure[]{\includegraphics[width=7.5cm]{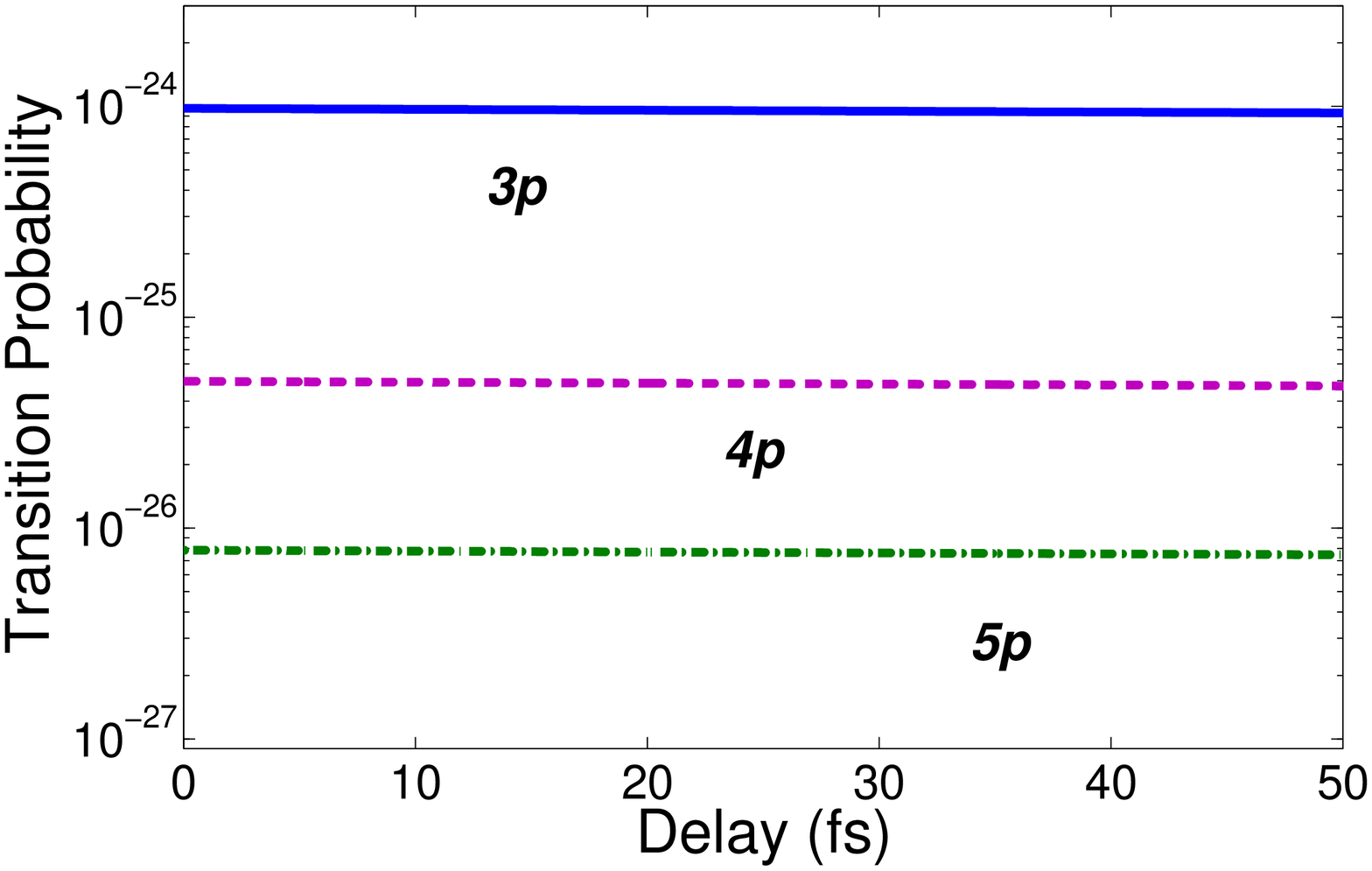}}
       \end{center}
\caption{Joint spectrum and single-intermediate level TPA transition probability as a
function of the delay $\tau$ for (a,b) Gaussian mode function, (c,d) sine
cardinal mode function and (e,f) uncorrelated photons. Intermediate levels correspond to: $3p$
(blue solid line), $4p$ (violet dashed line) and $5p$ (green dash-dotted line). Pump pulse duration is set
 to $T_{+}=10$ ps and $T_{-} = 2$ ps.}
\end{figure}

Provided that $T\gg 1 / \abs{\Delta^{\pare{j}} -
\Delta^{\pare{k}}}$, to eliminate unwanted terms at intermediate
frequencies, a straightforward Fourier analysis of Eq.
(\ref{weighted}) reveals the curve shown in Fig. 3. We see that
peaks emerge from the Fourier transform of the
weighted-and-averaged TPA transition probability, whose locations
determine the energy mismatch of the intermediate states: $5.1$,
$6.98$, $7.65$, $7.95$, and $8.12$ eV. With these values and the
definition of the energy mismatch, we obtain the virtual-state
energy values: $-3.40$, $-1.51$, $-0.85$, $-0.54$, and $-0.37$ eV.
These energy values can be readily identified with $n=2, 3, 4, 5,
6$ corresponding to the $2p$, $3p$, $4p$, $5p$, and $6p$ states,
respectively. In obtaining Fig. 3, we have computed the average over $T_{-}$ with a time step $\delta T_{-} = 3$ fs. However, one can obtain the same results using a larger time step (up to $60$ fs) to reduce (by an order of magnitude) the amount of experiments that are needed to calculate the weighted-and-averaged TPA transition probability.

To get a clearer picture that two-photon virtual-state
spectroscopy depends on the quantum interference from different
contributions of intermediate-state transitions with a specific
spectral shape, let us consider a simpler, even though ideal, case
where a single intermediate quantum state ($3p$, $4p$ or $5p$) is
present \cite{nakanishi2009}. Figure 4 shows the two-photon
transition probability as a function of the delay $\tau$ for a
fixed value of $T_{-}$ and $T_{+}$, considering three different
intermediate states. Notice that in the case of an entangled
two-photon state bearing a Gaussian mode function [Fig. 4(a,b)],
and an uncorrelated two-photon state [Fig. 4(e,f)], contributions
from different intermediate transitions are monotonically
dependent on the delay $\tau$. In contrast, when considering an
entangled two-photon state with a sine cardinal mode function
[Fig. 4(c,d)], contributions from different intermediate states
exhibit an oscillatory behavior, whose frequency of oscillation
corresponds precisely to the frequency of each transition. In
consequence, the coherent summation of these contributions [Eq.
(\ref{prob_sinc})] leads to nonmonotonic variations in the TPA
signal [Fig. (2)] that carry information about the frequency of
all intermediate-state transitions. This information can then be
extracted by means of a Fourier analysis of the
weighted-and-averaged TPA signal [Eq. (\ref{weighted})].

The physical reason why two-photon states with similar degree of entanglement, but
different spectral shape, give rise to such contrasting results comes from the fact that
TPA probabilities are significantly affected by the shape of the two-photon mode function, as it has been shown, for instance, in Ref. \cite{nakanishi2009}. By increasing the time difference between the absorbed photons, i.e., increasing $T_{-}$ or $\tau$, one would expect a monotonic decay of the TPA signal, which is precisely what is observed with
a Gaussian spectral shape. Surprisingly, when considering a sine cardinal spectral shape (rectangular in the time domain),
one can find values of $T_{-}$ and $\tau$ where TPA is no longer observed, a phenomenon called entanglement-induced two-photon
transparency \cite{fei1997}. In two-photon virtual-state spectroscopy, we benefit from this behavior to extract information about the energy level structure of the medium under study.

It is worth remarking that the particular choice of the pump
duration $T_{+}$ does not modify the presented results, since its
value does not affect the way in which contributions from
different intermediate levels interfere [see Eq.
(\ref{prob_sinc})]. Additionally, we highlight the fact that the
same information as the one depicted in Fig. 3 can also be
obtained when quasi-uncorrelated photons ($T_{-} = 40$ ps) are
used, meaning that virtual-state spectroscopy can also be
performed even with a low degree of entanglement between the
photons. This low degree of entanglement, however, results in a lower
TPA transition probability [see Eq. (\ref{prob_sinc})], which might affect the signal-to-noise
ratio of an experimentally measured TPA signal. This highlights the role of the particular spectral shape
of the paired photons used in two-photon virtual-state spectroscopy. While a proper spectral shape of the
photons guarantees a successful realization of this technique, the degree of entanglement controls the strength
of the TPA signal that is measured.

\section{Conclusions}

We have shown that virtual-state spectroscopy cannot be
performed by means of two uncorrelated rectangular-shaped
classical pulses, contrary to what it is suggested in Ref. \cite{mukamel2009}.
Also, we have shown that non-entangled frequency-correlated two-photon
states exhibit no dependence of the transition probability on the temporal
delay, so they are useless for performing virtual-state spectroscopy.
This implies that, in order to extract information about the
energy levels of a medium, one has to make use of two-photon
states bearing nonclassical frequency correlations. Interestingly, we have found
that more important than the degree of entanglement present, it is the specific
spectral shape of these correlations which allows one to perform two-photon
virtual-state spectroscopy. We have demonstrated that
while entangled states with a Gaussian spectral shape and a high
degree of entanglement cannot be used to perform virtual-state spectroscopy,
surprisingly, entangled two-photon states with a sine cardinal spectral shape and a very low degree
of entanglement can be used instead.

The results presented here help to identify clearly which types of two-photon sources
can be used to experimentally implement virtual-state
spectroscopy. By clarifying the role of entanglement, we have
found that even paired photons with a low degree of entanglement,
but with the appropriate sine cardinal spectral shape, guarantee
the successful realization of virtual-state spectroscopy. This
implies that entanglement by itself is not the key ingredient to
experimentally perform virtual state spectroscopy.

Finally, this work is also part of a greater research
effort devoted to identifying what physical effects necessarily require
the presence of entanglement to be observed.
Entanglement is a special type of correlation which exists between
two parties, i.e., two photons. However, photons can show
different types of correlations without entanglement. When a
certain effect is observed making use of entangled photons, it
might happen that this effect could also have been observed with
non-entangled photons, provided that the enabling factor is a specific
characteristic of the correlations that is shared between entangled and
non-entangled beams of photons.  Therefore, it becomes of
fundamental relevance to determine whether certain effects are due
to the existence of entanglement or to another accompanying
characteristic that can exist without its presence.

For instance, in sum-frequency generation (SFG), the
flux of generated photons increases with the
bandwidth of the incoming fundamental photons \cite{dayan_2005}. The bandwidth of the absorbed photons
can be made extremely large with appropriately engineered SPDC
sources \cite{hendrych2009}, which at the same time produces
entangled photons with an extremely large degree of entanglement.
However, the dependence of the flux rate on the bandwidth applies
as well to classical pulses. What it is unique to entangled
photons is the linear dependence of the rate on the number of
fundamental photons \cite{dayan_2005}. Paired photons produced in
SPDC can also be used to calibrate detectors \cite{migdall_2005}.
In this case, the key enabling factor is the presence of two
photons, since SPDC generates necessarily photons in pairs, but
not their frequency-entangled nature. When one photon is detected
and the other is not, we can infer that this is due to the
inefficiency of the detectors. Therefore, by taking the number of photons
detected in each detector, and the coincidence counts of paired photons
detected in both detectors, we are able to measure the efficiency of each
detector. Finally, several protocols proposed for spectroscopy
\cite{scarcelli_2003} also make use of frequency correlations
between photons rather than entanglement. This is closely related
to the demonstration of the possibility to use thermal (or
pseudothermal), and thus non-entangled, radiation for two-photon
imaging experiments \cite{valencia2005}. As it was demonstrated in
\cite{torres-company2011}, entangled and non-entangled sources can
show strikingly similar behaviors when traversing the same optical
system, characterized by a particular transfer function, provided that
certain properties of the frequency correlations between photons are the same for
both sources.

\ack

This work was supported by Government of Spain (Project
FIS2010-14831) and the European Union (FET-Open 255914,
PHORBITECH), and by the Fundacio Privada Cellex Barcelona.


\section*{References}
\begin{thebibliography}{XX}

\bibitem{denk1990} Denk W, Strickler J H, and Webb W W 1990 \emph{Science} \textbf{248} 73

\bibitem{hopfield1965} Hopfield J J and Worlock J M 1965 \emph{Phys. Rev.} \textbf{137} A1455

\bibitem{mukamel_book} Mukamel S 1995 \emph{Principles of Nonlinear Optical Spectroscopy} (Oxford University Press, New York)

\bibitem{mandel} Mandel L and Wolf E 1995 \emph{Optical Coherence and Quantum Optics} (Cambridge University Press, New York)

\bibitem{juha1990} Javanainen J and Gould P L 1990 \emph{Phys. Rev. A} \textbf{41} 5088

\bibitem{fei1997} Fei H -B, Jost B M, Popescu S, Saleh B E A and Teich M C 1997 \emph{Phys. Rev. Lett.} \textbf{78} 1679

\bibitem{teich1998} Saleh B E A, Jost B M, Fei H -B and Teich M C 1998 \emph{Phys. Rev. Lett.} \textbf{80} 3483

\bibitem{kojima2004} Kojima J and Nguyen Q -V 2004 \emph{Chem. Phys. Lett.} \textbf{396} 323

\bibitem{mukamel2012} Schlawin F, Dorfman K, Fingerhut B P and Mukamel S 2012 \emph{arXiv:1204.4490v1}

\bibitem{nasr2003} Nasr M B, Saleh B E A, Sergienko A V and Teich M C 2003 \emph{Phys. Rev. Lett.} \textbf{91} 083601

\bibitem{gouet2010} Le Gouët J, Venkatraman D, Wong F N C and Shapiro J H 2010 \emph{Opt. Lett.} \textbf{35} 1001

\bibitem{franson1992} Franson J D 1992 \emph{Phys. Rev. A} \textbf{45} 3126

\bibitem{gisin1998} Brendel J, Zbinden H and Gisin N 1998 \emph{Opt. Comm.} \textbf{151} 35

\bibitem{kim2009} Baek S Y, Cho Y W and Kim Y  H  2009 \emph{Opt. Express} \textbf{17} 19244

\bibitem{torres-company2011} Torres-Company V, Valencia A, Hendrych M and Torres J P 2011 \emph{Phys. Rev. A} \textbf{83} 023824

\bibitem{torres-company2012} Torres-Company V, Torres J P and Friberg A T 2012 \emph{Phys. Rev.
Lett.} \textbf{109} 243905

\bibitem{harris2008} Harris S E 2008 \emph{Phys. Rev. A} \textbf{78} 021807

\bibitem{harris2009} Sensarn S, Yin G Y and Harris S E 2009 \emph{Phys. Rev. Lett.} \textbf{103} 163601

\bibitem{dayan_2005} Dayan B, Pe'er A, Friesem A A and Silberberg Y 2005 \emph{Phys. Rev. Lett.} \textbf{94}, 043602

\bibitem{lee_2006} Lee D-I and Goodson III T 2006 \emph{J. Phys. Chem. Lett. B} \textbf{110}, 25582

\bibitem{lee_conf} Lee D-I and Goodson III T 2007 \emph{IEEE/LEOS Summer Topical Meetings, 2007 Digest of the,} 15-16

\bibitem{shore1979} Shore B W 1979 \emph{Am. J. Phys.} \textbf{47} 262

\bibitem{sakurai} Sakurai J J 1994 \emph{Modern Quantum Mechanics} (Addison-Wesley, USA)

\bibitem{mukamel2009} Roslyak O and Mukamel S 2009 \emph{Phys. Rev. A} \textbf{79} 063409

\bibitem{perina1998} Pe\v{r}ina Jr J, Saleh B E A and Teich M C 1998 \emph{Phys. Rev. A} \textbf{57} 3972

\bibitem{mollow1968} Mollow B R 1968 \emph{Phys. Rev.} \textbf{175} 1555

\bibitem{juan2011} Torres J P, Banaszek K and Walmsley I A 2011 \emph{Progress in Optics} \textbf{56} 227

\bibitem{law2000} Law C K, Walmsley I A and Eberly J H 2000 \emph{Phys. Rev. Lett.} \textbf{84} 5304

\bibitem{etherton1970} Etherton R C, Beyer L M, Maddox W E and Bridwell L B 1970 \emph{Phys. Rev. A} \textbf{2} 2177

\bibitem{cesar1996} Cesar C L, Fried D G, Killian T C, Polcyn A D, Sandberg J C, Yu I A, Greytak T J, Kleppner D and Doyle J M 1996 \emph{Phys. Rev. Lett.} \textbf{77} 255

\bibitem{book_bethe} Bethe H A and Salpeter E E 2008 \emph{Quantum Mechanics of One- and Two-Electron Atoms} (Dover Publications, New York)

\bibitem{bebb1966} Bebb H B and Gold A 1966 \emph{Phys. Rev.} \textbf{143} 1

\bibitem{contradiction} The origin of this contradiction lies on the use, in ref. \cite{mukamel2009}, of a wrong identity for multiplication of rectangular functions. This identity creates correlations between the fields, which ultimately lead to a nonmonotonic behavior of the TPA transition probability.

\bibitem{keller1997} Keller T E and Rubin M H 1997 \emph{Phys. Rev. A} \textbf{56} 1534

\bibitem{hendrych2007} Hendrych M, Micuda M and Torres J P 2007 \emph{Opt. lett.} \textbf{32} 2339

\bibitem{nakanishi2009} Nakanishi T, Kobayashi H, Sugiyama K and Kitano M 2009 \emph{J. Phys. Soc. Jpn.} \textbf{78} 104401

\bibitem{joobeur1994} Joobeur A, Saleh B E A and Teich M C 1994 \emph{Phys. Rev. A} \textbf{50} 3349

\bibitem{shih} Shih Y H and Sergienko A V 1994 \emph{Phys. Lett. A }\textbf{191} 201

\bibitem{hendrych2009} Hendrych M, Shi X, Valencia A and Torres J P 2009 \emph{Phys. Rev. A} \textbf{79} 023817

\bibitem{migdall_2005} Migdall A L, Datla R U, Sergienko A, Orszak J S and Shih Y H 1995 \emph{Metrologia} \textbf{32} 479

\bibitem{scarcelli_2003} Scarcelli G, Valencia A, Gompers S and Shih Y H 2003 \emph{Appl. Phys. Lett.} \textbf{83} 5560

\bibitem{valencia2005} Valencia A, Scarcelli G, D'Angelo M and Shih Y 2005 \emph{Phys. Rev. Lett.} \textbf{94} 063601



\end{thebibliography}
\end{document}